# Micro Power Supply Based on Piezoelectric Effect

M.Husak, A.Budkova, T.Pycha, A.Laposa, V.Povolny, V.Janicek, J.Novak, A.Boura, J.Foit

Department of Microelectronics, Faculty of Electrical Engineering, CTU in Prague
Technicka 2, 166 27 Prague 6, Czech Republic
e-mail: husak@fel.cvut.cz

*The paper solves the model of the miniature Power supply based on the piezoelectric cantilever. The aim of the future is to further hybrid integration and use of nanotechnology. Contents of the article belongs to the category of renewable energy sources with environment energy conversion into electrical energy. The work is focused on the use in small temperature differences.*

## 1. Introduction

The information contained in the article are based on known principles of the piezoelectric effect. This is a follow-up work with the output results achieved in 2022. The aim was to verify the possibility of using piezoelectric effect, verifying properties using model, determine the essential characteristics, finding the optimum load, output voltage and output power achieved. Attention is given to the basic principles of activities, electronic circuit connection as well as behavior of the piezoelectric cantilever. The attention is focused on information about the electrical design, information of the measured parameters, where have been achieved the best results.

## 2. Tested piezoelectric generators

Two types of piezoelectric elements were tested (Mide). The first sample is the piezoelectric cantilever element S128-H5FR-1107YB (Mide), the element was used was used to implement the model. It is a single-layer beam with piezo element made of ceramic material PZT 5H sealed in fiberglass epoxy laminate FR4. The thickness of the whole cantilever is 0.71mm, its dimensions are 53.0mm x 20.8mm. The dimensions of the piezo element are 27.8mm x 18.0mm x 0.19mm. The weight of the beam is 2 grams (2g). Cu electrodes for charge collection are located on the bottom and top of the piezo element. The second tested sample labeled S233-H5FR-1107XB has a bimorph structure (double layer, i.e. two layers of 5H piezoelectric ceramics).

The measuring station was designed for testing the real parameters of piezoelectric modules. The workplace is made up of a vibrating table controlled by an excitation generator, an optical distance sensor, an oscilloscope, weight scales and a resistive load decade - Fig. 1.

An external optical radiation source PS2520G (Tektronix) powers the IFS2401-1 sensor via a DT 2451 controller. The IFS2401-1 probe projects polychromatic light (white light) onto the target surface to be measured. The sensor lenses are designed to use controlled chromatic aberration to focus each wavelength of light to a specific distance. The light reflected from the target surface passes through the confocal aperture into the spectrometer, which detects and processes the spectral changes. The acquired data is transferred to the controller. The data from the spectral analysis of the light is used in the controller to calculate

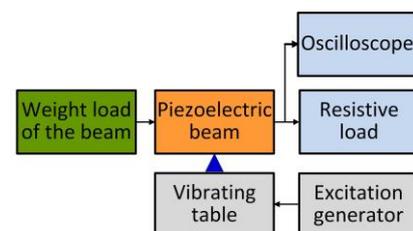

Fig. 1 Block diagram of the measuring workplace

## 3. Control of the piezoelectric generator

The operation control of the piezoelectric generator was realized with the circuit LTC3588-1. The circuit integrates a low-loss full-wave bridge rectifier with a high efficiency buck converter to form a complete energy harvesting solution optimized for high output impedance energy sources such as piezoelectric transducer. An ultralow quiescent current undervoltage lockout (UVLO) mode with a wide hysteresis window allows charge to accumulate on an input capacitor until the buck converter can efficiently transfer a portion of the stored charge to the output. The buck converter turns on and off as needed to maintain regulation. Four output voltages, 1.8V, 2.5V, 3.3V and 3.6V, are selectable with up to 100mA of continuous output current. Input operating range of the circuit is 2.7V to 20V Typical applications of the circuit are piezoelectric Energy Harvesting, electro-mechanical energy harvesting, wireless HVAC sensors, mobile asset tracking, tire pressure sensors, battery replacement for Industrial sensors, remote light switches, standalone nanopower buck regulators. The electrical connection of the LTC3588-1 for evaluating the output electrical energy from the piezoelectric transducer will be in the full paper, the block connection of the measuring system of the source model is shown in Fig. 2.

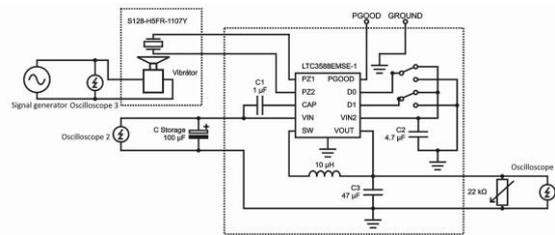

Fig. 2 Block connection of the measuring system of the source model

The output signal from the piezoelectric transducer is alternating. The signal must be rectified, stabilization of the output supply voltage must be ensured. An integrated circuit (IC) from Linear Technology has been selected for this purpose. The IC is primarily intended for controlling the collection of electrical energy generated by vibrations, it is designed for operation with high efficiency and for an output voltage of 1.8 to 3.6V and a current of up to 100μA. The IC consists of three basic blocks (bridge rectifier, UVLO control block and step-down DC/DC converter). The operation control of the piezoelectric generator was realized with the circuit LTC3588-1. The circuit integrates a low-loss full-wave bridge rectifier with a high efficiency buck converter to form a complete energy harvesting solution optimized for high output impedance energy sources such as piezoelectric transducer. An ultralow quiescent current undervoltage lockout (UVLO) mode with a wide hysteresis window allows charge to accumulate on an input capacitor until the buck converter can efficiently transfer a portion of the stored charge to the output.

## 4. Achieved results

Basic measurements of properties of 2 types of beams were performed. The measurements were carried out with the electrical output empty and with a resistive load, with different load weights at the end of the beams. Measurements were made for a range of frequencies from 16Hz to 500Hz. The measured range was extended by frequencies: 170Hz, 175Hz, 180Hz, 190Hz, 195Hz, 200Hz for a more accurate determination of the resonance frequency. The ends of the beams were loaded with 0g, 0.6g, 1.0g, 1.5g during measurement. The measured range of frequencies was expanded for a more accurate determination of the resonant frequency in a given configuration of the measuring set. For example, the S128-H5FR-1107YB parameters were measured at a load of 0.6 g and for frequencies: from 110Hz to 150Hz. The maximum measured voltage on the S128-H5FR-107YB beam was 26.9V at a mass load of 1.0g. The S233-H5FR-107XB beam generated the maximum voltage at the maximum mass load used of 1.5g. The bimorph beam at the resonance frequency generated a voltage higher than a unimorphic beam. The resonant frequency (at which the output voltage is maximum) is dependent on the weight of the load at the free end of the beam. For example, for S128-H5FR-1107YB at a mass load of 1.0g, the resonant frequency is 100Hz. For the same amplitude but a mass load of 1.5g, the resonant frequency is 90Hz. An example of measuring the dependence of voltage on frequency for S233-H5FR-1107XB at different mass loads is shown in Fig. 3. The resonant frequency decreases with increasing mass loading of the end of the beam, the generated voltage increases with increasing mass loading.

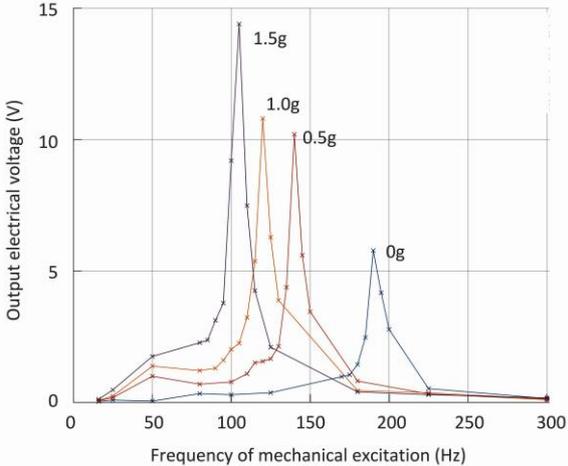

Fig. 3 Dependence of electrical voltage on frequency of mechanical excitation for S233-H5FR-1107XB at different weight loads at the end of the beam

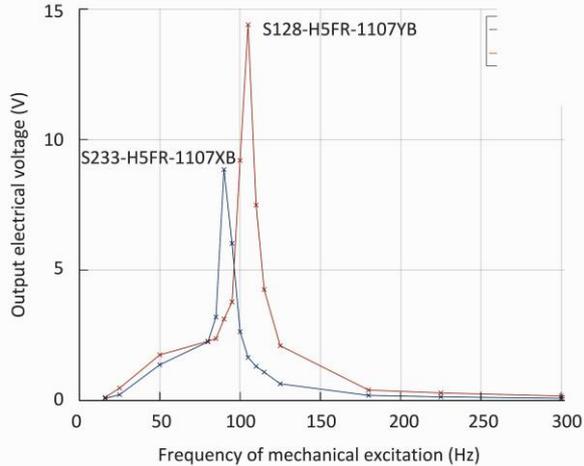

Fig. 4 Comparison of electrical voltage versus frequency for S128-H5FR-1107YB and S233-H5FR-1107XB at 1.5g mass load

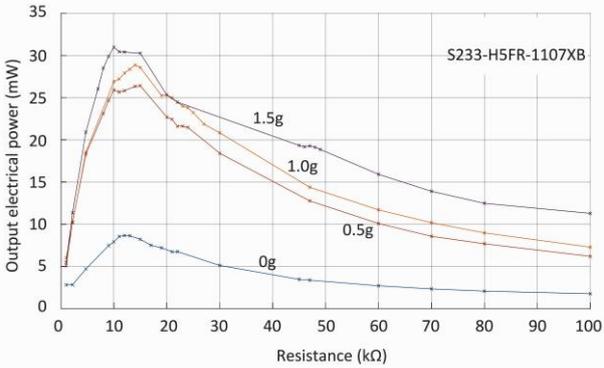

Fig. 5 The dependence of the output electrical power of S233-H5FR-1107YB on the resistive load at different mass loads

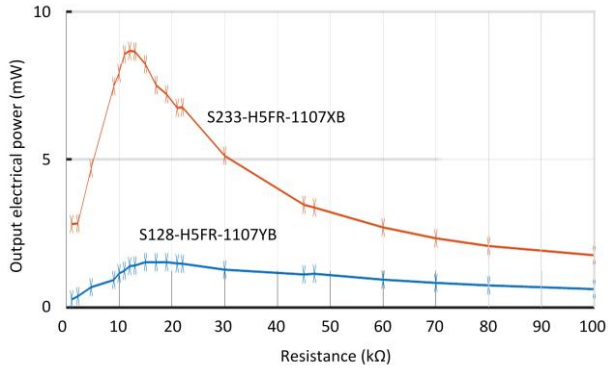

Fig. 6 Comparison of the output powers of the piezoelectric modules S128-H5FR-1107YB and S233-H5FR-1107XB depending on the resistive loads

The comparison of piezoelectric modules S128-H5FR-1107YB and S233-H5FR-1107XB under a mass load of 1.5g is shown in Fig. 4. The generated electric voltage of the bimorph module S233-H5FR-1107XB is greater than the electric voltage of the unimorph module S128-H5FR–1107YB. The dependence of the output electrical power of S233-H5FR-1107YB on the resistive load at different mass loads of the free end of the beam is shown in Fig. 5. Fig. 6 presents a comparison of the achieved output powers of the piezoelectric modules S128-H5FR-1107YB and S233-H5FR-1107XB depending on the resistive load at a constant mass load.

After the final implementation of the piezoelectric generator with control circuits, the output power achieved is the verification of the resonant frequency - Fig. 7. Measured resonant frequency for 3 loads (4.7kΩ, 10kΩ and 22kΩ) with a frequency change step of 2Hz. The resonant frequency corresponds to about 176Hz. The flattening of the load curve $R_{load}$ = 22kHz is given by the limited maximum value of the integrated circuit output voltage. The value was set to 3.6V for this measurement.

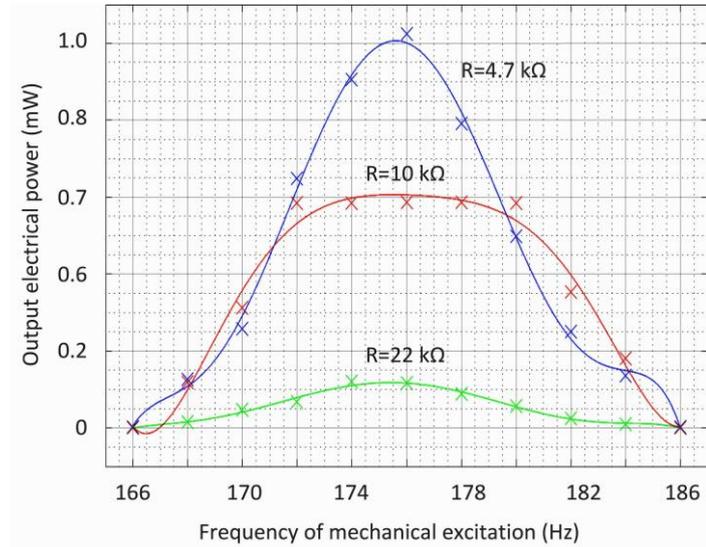

Fig. 7 Electrical output power with verification of the resonant frequency

## 5. Conclusions

The parameters of the two piezoelectric elements were measured, the output voltage, the output power depending on the frequency and the load on the free end of the beams. A power supply with a power management control circuit and piezoelectric beams was realized. The output parameters of the power supply were measured using a supercapacitor as energy storage.

**Acknowledgement**

This research has been supported by the EU project Horizont "GaN-for-Advanced-Power", No. 101007310 — GaN4AP — H2020-ECSEL-2020-1-IA-two-stage, and partially by CTU project No. SGS20/175/OHK3/3T/13 Integrated and phonic circuits and microstructures and the project "Research center for data analysis and protection - II. Stage (SmartLife)".